\documentclass{WileyMSP-template}

\usepackage{amsmath}
\usepackage{graphicx}
\usepackage{xcolor}
\usepackage{amsthm}
\usepackage{subfigure}
\usepackage{csquotes}
\usepackage{changes}

\newcommand{\bra}[1]{\langle #1|}
\newcommand{\ket}[1]{|#1\rangle}

\begin{document}


\title{A many-body perturbation theory approach to energy band alignment at the crystalline tetracene-silicon interface}

\maketitle


\author{Mykhailo~V.~Klymenko}
\author{Liang~Z.~Tan} 
\author{Salvy~P.~Russo}
\author{Jared~H.~Cole}

\begin{affiliations}
M.~V.~Klymenko, S.~P.~Russo, and J.~H.~Cole\\
ARC Centre of Excellence in Exciton Science, School of Science, RMIT University, Melbourne, Victoria 3001, Australia\\
Email Address: mike.klymenko@rmit.edu.au

L.~Z.~Tan \\
The Molecular Foundry, Lawrence Berkeley National Laboratory, Berkeley, California 94720, United States
\end{affiliations}


\keywords{hybrid inorganic-organic semiconductor interface, band edges, energy band alignment, GW approximation}

\begin{abstract}
    Hybrid inorganic-organic semiconductor interfaces are of interest for new photovoltaic devices operating above the Shockley-Queisser limit. Predicting energy band alignment at the interfaces is crucial for their design, but represents a challenging problem due to the large scales of the system, the energy precision required and a wide range of physical phenomena that occur at the interface. To tackle this problem, we use many-body perturbation theory in the non-self-consistent GW approximation, orbital relaxation corrections for organic semiconductors, and line-up potential method for inorganic semiconductors which allows for tractable and accurate computing of energy band alignment in crystalline van-der-Waals hybrid inorganic-organic semiconductor interfaces. In this work, we study crystalline tetracene physisorbed on the clean hydrogen-passivated 1x2 reconstructed (100) silicon surface. Using this computational approach, we find that the energy band alignment is determined by an interplay of the mutual dynamic dielectric screening of two materials and the formation of a dipole layer due to a weak hybridization of atomic/molecular orbitals at the interface. We also emphasize the significant role of the exchange-correlation effects in predicting band offsets for the hybrid inorganic-organic semiconductor interfaces.
\end{abstract}


\section{Introduction}

The interest in crystalline hybrid inorganic-organic semiconductor (HIOS) heterostructures is motivated by their potential to increase the internal quantum efficiency of photovoltaic devices \cite{MacQueen}, overcoming the Shockley-Queisser limit \cite{Shockley-Queisser}. In many modern proposals of HIOS devices, the organic layer facilitates efficient optical generation of excitons via the singlet fission effect, while the inorganic layer provides efficient separation and transport of charge carriers towards the electrodes \cite{MacQueen,Einzinger2019,Niederhausen2020}. Such device operation requires efficient energy transfer across the interface \cite{schlesinger2016energy, zhou2021optoelectronic}. Recent experimental characterizations of the interface between hydrogen passivated silicon (Si) and tetracene (Tc) have shown rather weak energy transfer for the triplet excitons \cite{MacQueen,Einzinger2019}. However, it has been also recently demonstrated that introducing HfON inorganic inter-layer between Si and Tc can improve the transfer \cite{Einzinger2019}. One of the obstacles preventing efficient energy transfer is exciton dissociation at the interface. This process can involve transitions through a series of the inter-layer charge transfer exciton states \cite{Bakulin2013}, which, in turn, depend on the energy band alignment. It is therefore crucial to have an accurate understanding of the band alignment in the vicinity of the interface.

 \begin{figure}[!t]
    \centering
    \includegraphics[width=8.0 cm]{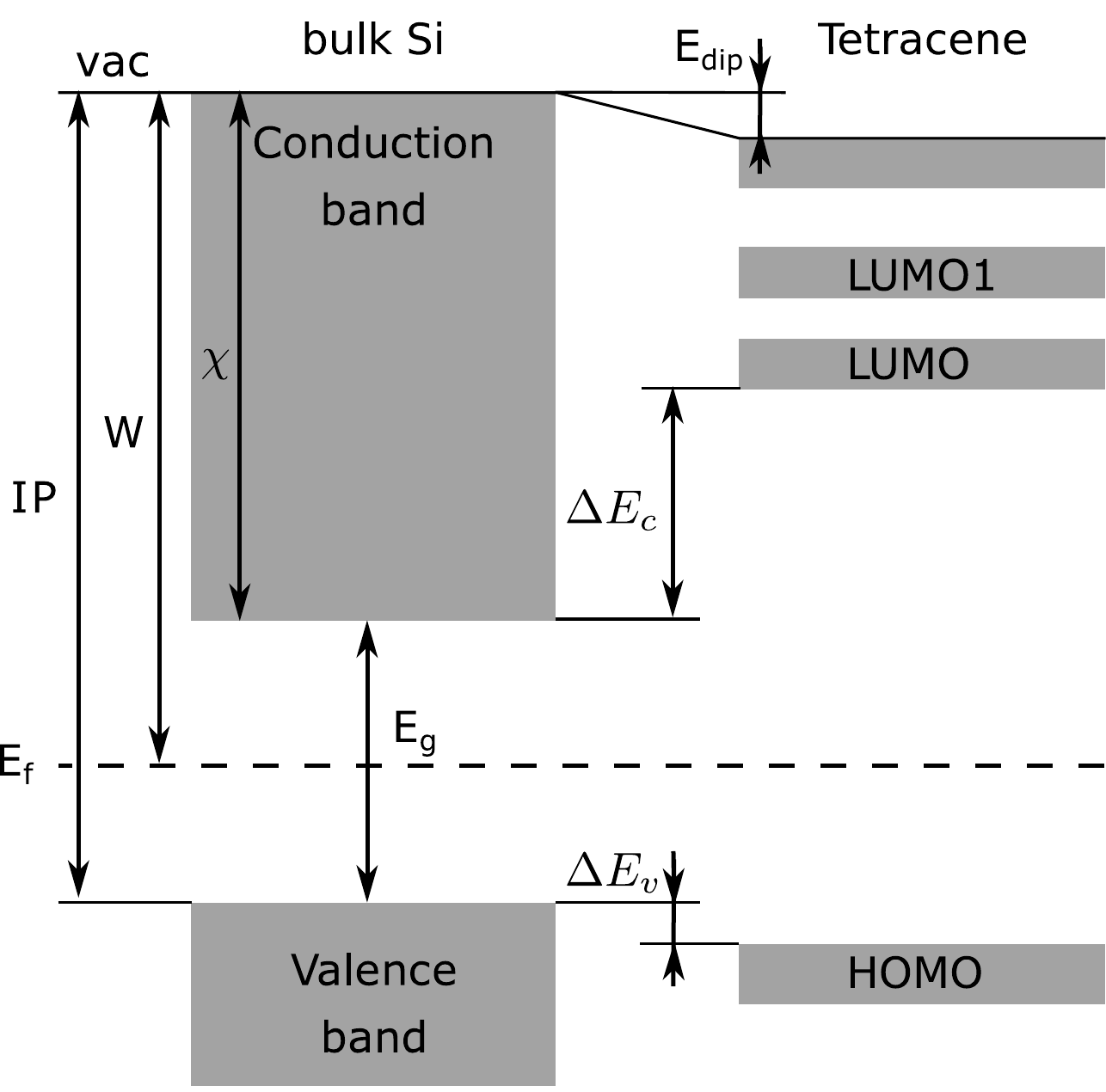}
    \caption{Alignment of the energy bands at the interface between two semiconductors determines the band offsets, $\Delta E_v$ and $\Delta E_c$. The energy band alignment is characterized by the ionization potential (IP), electron affinity ($\chi$), band gap ($E_g$) and work function (W). For the HIOS, the interface slightly modifies the energy bands due to mutual dynamic dielectric screening and formation of an interfacial dipole layer ($E_{dip}$).}
    \label{fig:b_diag}
\end{figure}

A typical band diagram of the HIOS interface is shown in Fig. \ref{fig:b_diag}. The widths of energy bands in the organic semiconductors are usually smaller than in the inorganic ones; the former are characterized by a much weaker hybridization of molecular orbitals. Energy band alignment can be quantitatively characterized by band offsets, $\Delta E_v$ and $\Delta E_c$, which can be estimated if ionization potentials, $IP$, electron affinities, $\chi$, and the dipole layer energy, $E_{dip}$, are known. The experimental values of these quantities for the Si-Tc heterostructure have been reported in several works \cite{MacQueen, Einzinger2019, Niederhausen2020}.

There are many theoretical studies of energy levels alignment for a single molecule physisorbed or chemisorbed at the surface of a metal or inorganic semiconductor \cite{Neaton1,Neaton2,Neaton3,D0CP06605B,Hofmann_2013}. However, their predictive power can not be transfer directly to the HIOS interfaces since they don't take into account delocalization of charge carriers and dielectric screening in the media that takes place in organic crystals, especially with dense packing. On the other hand, many works on the organic crystalline materials are focused on the macroscopic electrostatic effects such as Fermi level pinning and band bending\cite{schlesinger2016energy} and are lacking information on the mutual dynamic dielectric screening and electron correlations across the interface \cite{Neaton2}.

One way to take these effects into account is by using density functional theory (DFT) with hybrid range-separated exchange-correlation functionals.\cite{Neaton1} This method is quite efficient in terms of computational resources, but on the other hand it has a free parameter that has to be derived either empirically or via additional first-principle computations. The first-principle computations of this parameter for HIOS interfaces faces some uncertainty \cite{Wruss2018} and cannot reproduce dynamical screening that can limit the accuracy in some cases.\cite{Neaton2} For the Si-Tc interface considered in this work, the recent attempt to compute the energy bands alignment using all-electron DFT with hybrid exchange-correlation functionals (see Ref. \cite{Janke_2020}) results in the so-called staggered gap heterojunction (type-II), while the experimental data shows that this interface must be the straddling gap heterojunction (type-I) according to Refs.  \cite{MacQueen} and \cite{Einzinger2019}. The definitions of the type-I and type-II heterojunctions can be found in Ref. \cite{ihn2010semiconductor}.

In inorganic semiconductor heterostructures with a small lattice mismatch, approximate values for the band offsets can be obtained using Anderson's rule \cite{davies_1997}. Usually, an inaccuracy of Anderson's rule is associated with the dipole layer that causes an energy shift, $E_{dip}$, of the vacuum level across the interface. For HIOS, additionally to the dipole layer, Anderson's law can be also violated due to a substantial difference in the nature of electron correlations in the organic and inorganic semiconductors.
 
Energies of the band edges are the quantities related to many-body excited states which can be rigorously expressed in terms of quasi-particle energies in the framework of many-body perturbation theory (MBPT) \cite{Onida, Golze}. In the case of inorganic semiconductors, however, the band edges can be obtained from DFT computations using the method proposed by Van-de-Walle and Martin \cite{van-de-walle}. According to this method, DFT is used to compute the positional dependence of the microscopically averaged electrostatic potential (or vacuum level) taking into account bend bending and dipole layer at the interface. Equipped with the computed electrostatic potential, one can align the band edges computed or measured for bulk semiconductors relatively to this potential and get the correct band offsets assuming that the difference in the exchange-correlation effects at each side of the heterojunction is negligibly small. 

The goal of this work is to estimate the role of the dipole layer, exchange-correlation effects and dielectric screening at the crystalline HIOS interface represented by the Tc/Si heterostructure. We  primary focus on the physisorbed molecules with dense coverage requiring large unit cells. For such structures, standard GW methods that work well for a single chemisorbed molecule such as in Ref. \cite{Turkina} perform poorly for the interfaces. In this work, we present the results of a series of numerical experiments aimed to predict energy band alignment at HIOS interfaces with physisorption using DFT and MBPT with the G$_0$W$_0$ approximation. Using DFT, we analyze the formation of the dipole layer and its nature in the Tc/Si heterostructure. Comparing results obtained by DFT and MBPT, we estimate the role of the exchange-correlation effects in the energy band alignment at the HIOS interface. Next, using the dielectric embedding technique \cite{Liu2019, Liu2020}, we isolate the effect of the dielectric screening from the effect of the dipole layer in order to estimate the contribution of each of them separately.

\section{DFT results for the supercell}

\subsection{Atomic structure of the Tc/Si interface and DFT results}

It has been experimentally observed that Tc can be deposited onto neutral surfaces in either so-called \enquote{flat-lying}  or \enquote{upright-standing} configurations \cite{Betti2007, Niederhausen2020}. In the first case, the angle between the long molecular axis and surface is small, while in the second case this angle is close to 90$^\circ$. In this work we are interested in the latter case, since it is characterized by a denser coverage of organic molecules \cite{Jaeckel2007} and, therefore, one may expect higher mobility of charge carriers in Tc, which is a desirable property for electronic applications.

\begin{figure}[!t]
    \centering
    \subfigure[]{\includegraphics[width=8.1 cm]{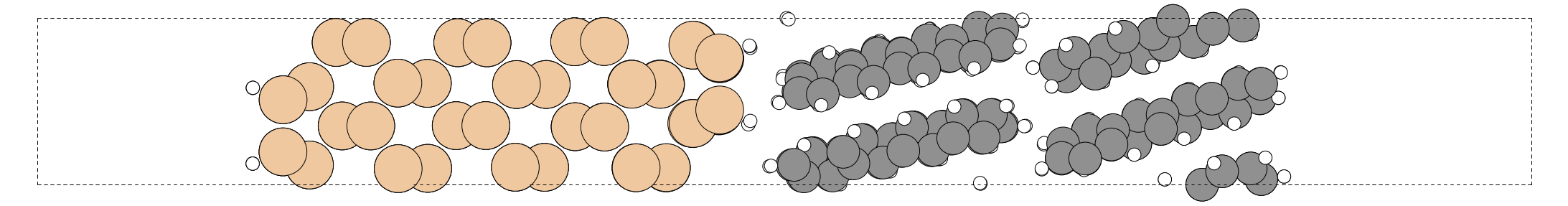}}\\
    \subfigure[]{\includegraphics[width=8.1 cm]{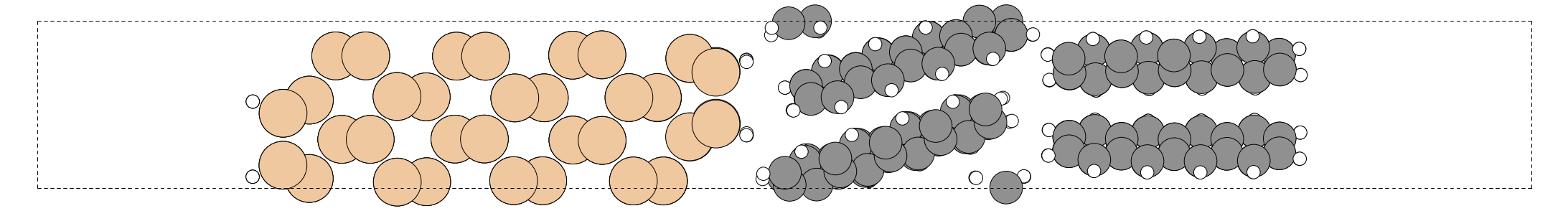}}\\
    \subfigure[]{\includegraphics[width=8.1 cm]{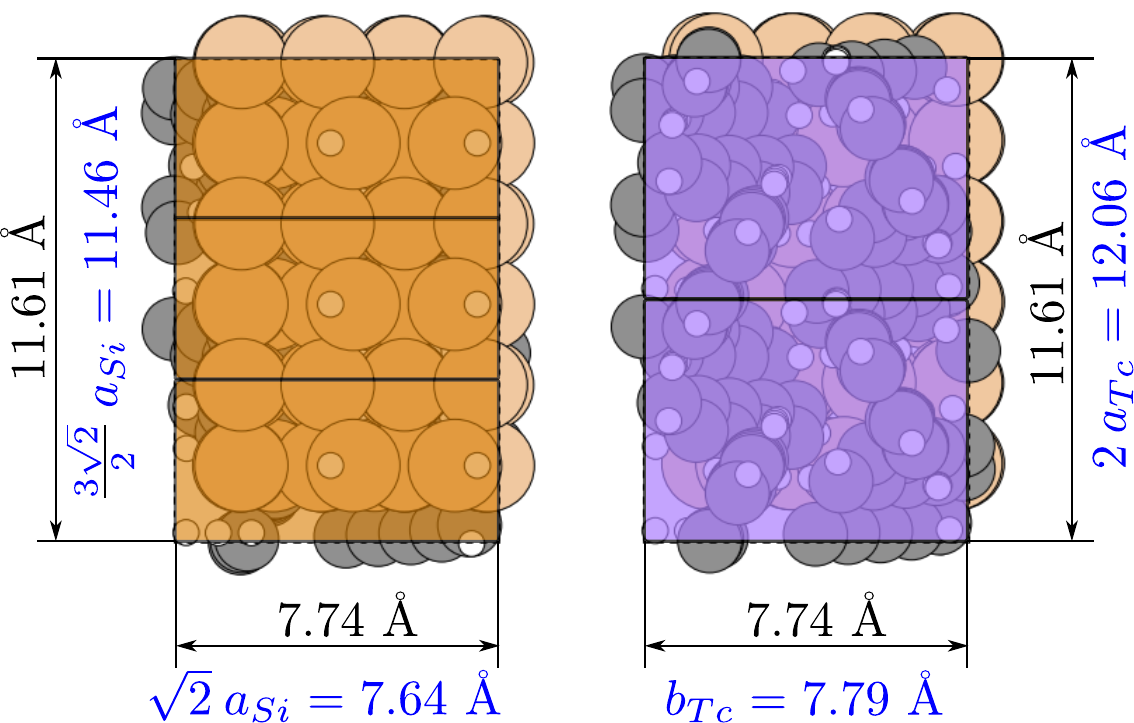}}
    \caption{Atomic structure of the van-der-Waals interface between the crystalline Si with the 1x2 reconstructed (100) surface and Tc a) before and b) after atomic relaxation. c) Alignment of the Si and Tc unit cells at the interface. Numbers in blue color ($a_{Si}$, $a_{Tc}$, $b_{Tc}$) denote sizes of unit cells for bulk materials published in the literature, black color numbers correspond to the actual sizes of the super-cell after relaxation computed in this work.}
    \label{fig:slab}
\end{figure}

We focus on one particular Tc/Si heterostructure with a relatively small crystal lattice mismatch that has been also thoroughly characterized experimentally \cite{PhysRevB.74.205326} - the contact between the \enquote{upright-standing}-configuration surface of the crystalline Tc and hydrogen-passivated 1x2 reconstructed (100) Si surface, shown in Fig. \ref{fig:slab}.

All computations in this work are performed for a supercell containing 16 atomic layers of Si and two molecular layers of Tc. We need at least two   molecular layers to simulate accurately the electronic environment for the Tc molecules at the interface. A single layer would be the surface layer \cite{Morisaki2014}.

The atomic coordinates have been obtained from the geometry optimization within DFT with a plane-wave basis set and norm-conservative pseudo-potentials \cite{Hamann, PseudoDojo}. The computations were performed using Quantum Espresso, the plane-wave DFT software \cite{Giannozzi_2017, QE}. The dispersion forces, responsible for the physisorption of Tc on the Si surface, are introduced in the model via the non-local exchange-correlation functional vdW-DF2-C09 \cite{C09}. The computations have been performed on the 4x3x1 Monkhorst-Pack k-space grid, using a kinetic energy cutoff of 80 Ry for wavefunctions and 320 Ry for charge densities. These choices are justified by a series of convergence tests (see Supplementary materials, Section 1). For the geometry optimization, we used the use Broyden-Fletcher-Goldfarb-Shanno quasi-newton algorithm with variable cell parameters. The system is periodic only in two in-plane dimensions. The effect of the periodic boundary conditions in the third dimension was canceled by the dipole correction \cite{PhysRevB.59.12301}.

The 1x2 reconstructed Si(100) surface has a rectangular unit cell with lateral sizes of $\sqrt{2} a_{Si}$ and $(\sqrt{2} / 2)  a_{Si} $ along the axes $[110]$ and $[\bar{1}10]$ respectively, where $a_{Si}$ is the lattice constant of bulk Si. The minimal unit cell spans giving the best match of sublattices are $1 \times 2$ for Tc and $1 \times 3$ for Si, shown schematically in Fig. \ref{fig:slab}. In order to estimate the mismatch, we take the lattice constants reported in the literature $a_{Si} = 5.4$ \AA\ for Si and $a_{Tc}= 7.9$ \AA\ and $b_{Tc}= 6.03$ \AA\  for Tc, then $\sqrt{2} a_{Si} = 7.64$ \AA\ and  $3 \times  (\sqrt{2} / 2)  a_{Si} = 11.462$ \AA\ that makes the mismatches along each of the axes being $0.26$ \AA\ and  $0.6$ \AA. 

\begin{figure}[!t]
    \centering
    \includegraphics[width=8.5 cm]{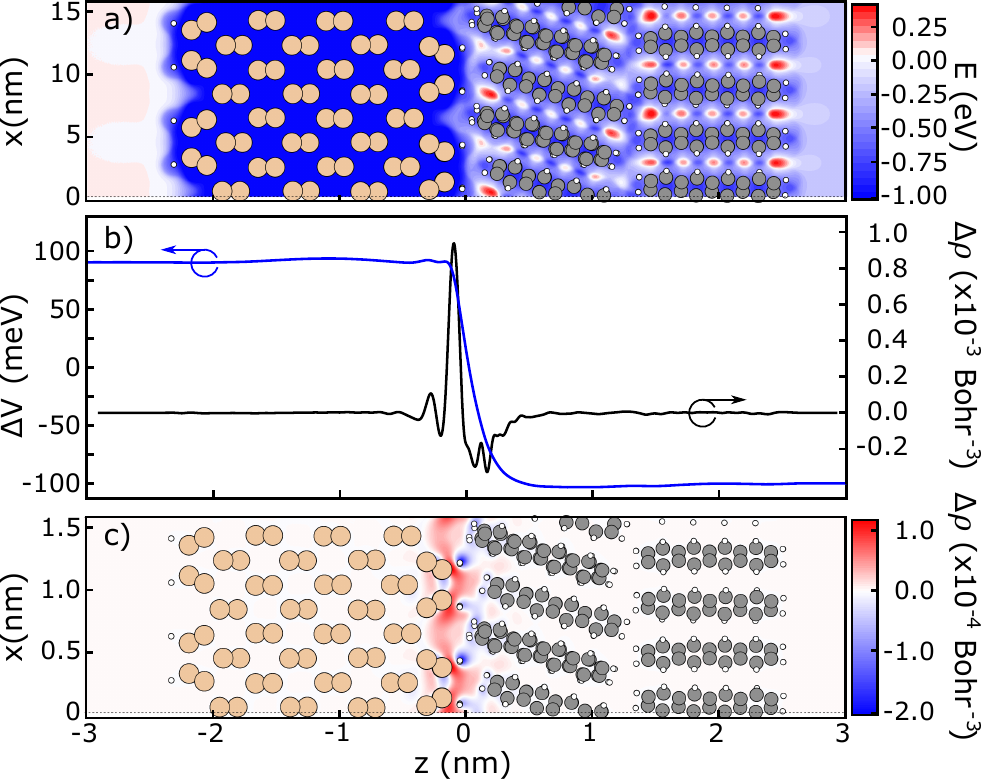}
    \caption{    \label{fig:si_tc_bands2} a) Isocontour map of the electrostatic potential in the vicinity of the vacuum level for the  the Tc/Si interface, b) in-plane averaged change of the electrostatic potential and electron density in the Tc/Si slab relative to isolated Si and Tc slabs, c) 2D plot of the electron density change (same as above but averaged in the y-direction only).}
\end{figure}

 The computed lateral sizes of the unit cell are 7.74 \AA\ and 11.61 \AA. These values are close to the experimentally observed lattice  constants  published in Ref. \cite{PhysRevB.74.205326} for the crystalline Tc deposed on the 1x2 reconstructed (100) Si surface: $a_{Tc}= 7.3 \pm 0.6$ \AA\ and $2b_{Tc}= 2 \times 5.5 \pm 0.6 = 11.0 \pm 0.6$ \AA. Comparing to the results for isolated slabs, we find that the Si slab remains almost undeformed imposing slight deformations on the Tc lattice. This can be explained by the fact that, comparing to deformations in Si, deformations of Tc require less energy due to the van-der-Waals nature of bonding in the molecular crystal. 
 
 The results of the geometry optimization also show that the most dramatic changes occur in the surface Tc layer, which agrees with the results of Ref. \cite{Morisaki2014}.

\subsection{Dipole layer}

Using DFT with the parameters described above, we have computed the electrostatic potential and charge density in the Si/Tc super-cell. The isocontour map of the electrostatic potential for energies close to the vacuum level of silicon is shown in Fig. \ref{fig:si_tc_bands2}a. On this map, the potentials to the left and to the right of the heterostructure are not equal which is indicative of the dipole layer. To estimate the mutual effect that Si and Tc layers make on each other, it is useful to compare the results for the heterostructure with the results for free surfaces given the atomic configuration remains unchanged. Qualitatively, this mutual effect can be expressed by the change of the in-plane averaged electrostatic potential and charge density (see Fig. \ref{fig:si_tc_bands2}b):

\begin{figure}[!t]
    \centering
    \includegraphics[width=8.5 cm]{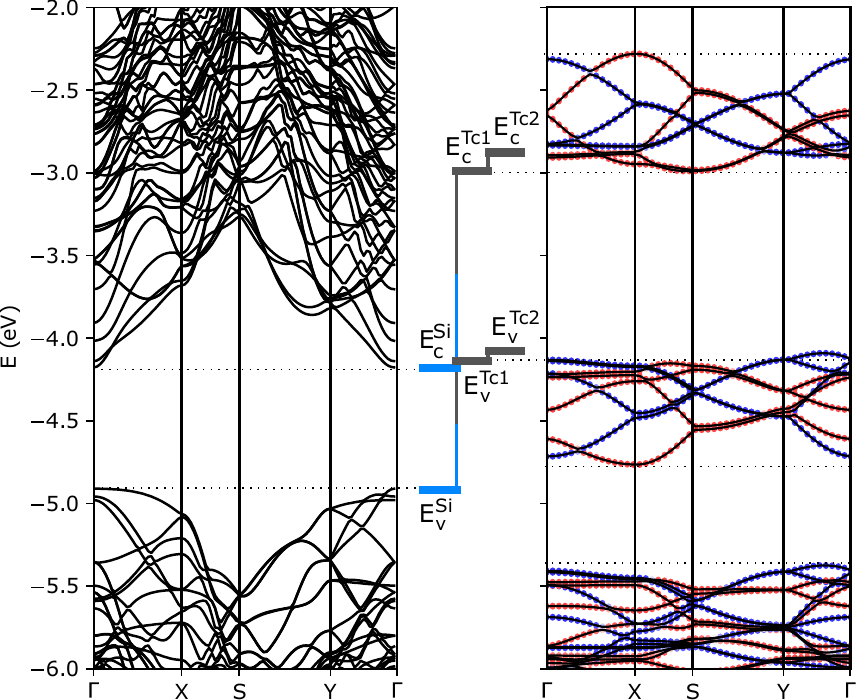}
    \caption{    \label{fig:si_tc_bands1} Band structure of the isolated Si and Tc slabs computed within DFT/LDA. The blue and grey bars denote the valance and conduction band edges in Si and Tc respectively.}
\end{figure}

\begin{equation}
\Delta V = V_{Si/Tc} - V_{Si} - V_{Tc},
\label{pot}
\end{equation}
\begin{equation}
\Delta \rho = \rho_{Si/Tc} - \rho_{Si} - \rho_{Tc},
\label{den}
\end{equation}
where the index $Si/Tc$ corresponds to the computations in the super-cell, and the indices $Si$ and $Tc$ correspond to the quantities computed for the isolated Si and Tc slabs respectively.

A step-wise change of the electrostatic potential and corresponding change of the charge density shown in Fig. \ref{fig:si_tc_bands2}b also evidence the appearance of the interfacial dipole layer with energy $E_{dip}=191$ meV. The energy $E_{dip}$ is defined as the difference between vacuum levels at the left and right sides of the slab. The formation of such a dipole layer requires additional interpretation since it can be caused by different reasons - physical effects or inherent errors of the used method. Possible systematic errors are related to the fact that DFT underestimates the bandgap which can lead to qualitatively incorrect energy band alignment that can result in a spurious charge transfer. Indeed, considering the band structure of isolated Si and Tc slabs (see Fig. \ref{fig:si_tc_bands1}), we observe that the maximum of the valence band in Tc is located higher than the minimum of the conduction band in Si making possible the electron tunneling from the occupied states in Tc to the unoccupied states in Si. Such an energy band alignment defines a type-II heterostructure and is prone to a spurious charge transfer resulting in a dipole layer. On the other hand, the dipole layer can be physical, resulted from a weak hybridization of atomic orbitals across the interface.

\begin{figure}[!t]
    \centering
    \includegraphics[width=7.0 cm]{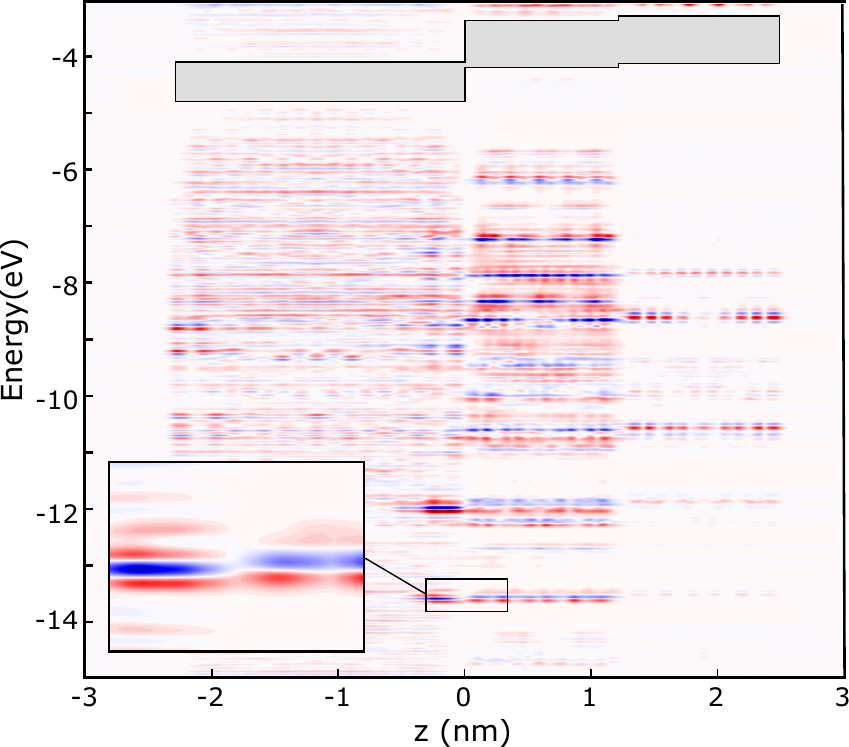}
    \caption{    \label{fig:ldos} In-plane averaged change of the local density of states in the Tc/Si slab relative to isolated Si and Tc slabs computed via Eq. (\ref{den1}). The grey shaded area indicates the band gaps of semiconductors. The local density of states change has features,  shown in the in-plane image, that are characteristic of the orbital hybridization with clear signatures of bound and anti-bound states.}
\end{figure}

In order to determine whether appearance of the dipole layer is caused by a spurious charge transfer or hybridization of atomic orbitals across the interface, we express the charge density difference in Eq. (\ref{den}) in terms of the local density of states (LDOS):

\begin{equation}
\begin{split}
\Delta \rho = & \int_{-\infty}^{E_f^{Si/Tc}} d\varepsilon LDOS_{Si/Tc}(\varepsilon) - \int_{-\infty}^{E_f^{Si}} d\varepsilon LDOS_{Si}(\varepsilon) - \int_{-\infty}^{E_f^{Tc}} d\varepsilon LDOS_{Tc}(\varepsilon) = \\
&\int_{-\infty}^{E_f^{Si/Tc}} d\varepsilon \left[ LDOS_{Si/Tc}(\varepsilon) - LDOS_{Si}(\varepsilon - E_f^{Si/Tc} + E_f^{Si}) - LDOS_{Tc}(\varepsilon - E_f^{Si/Tc} + E_f^{Tc}) \right] = \\
&\int_{-\infty}^{E_f^{Si/Tc}} d\varepsilon \Delta LDOS(\varepsilon),
\end{split}
\label{den1}
\end{equation}

where $LDOS_{Si/Tc}(\varepsilon)$, $LDOS_{Si}(\varepsilon)$ and $LDOS_{Tc}(\varepsilon)$ are local density of states for the Si/Tc supercell, Si and Tc isolated slab respectively; $E_f^{Si/Tc}$, $E_f^{Si}$ and $E_f^{Tc}$ are Fermi levels in the Si/Tc supercell, Si and Tc isolated slab respectively (the Fermi levels are pinned to the top of the valence band in the corresponding structures). LDOS in Eq. (\ref{den1}) is integrated in volumes defined by thin slices of the unit cell in the direction perpendicular to the interface, so the LDOS is a function of energy, $\varepsilon$, and one of the spatial coordinates, $z$ (see Fig. \ref{fig:si_tc_bands2}c). At the second line of Eq. (\ref{den1}), we align LDOS functions on the energy axis in order to apply a common integration domain for all the terms. This has been achieved by aligning the Fermi levels, assuming that the Fermi level is located at the top of the valence band.

The term in the square brackets is the change of LDOS, $\Delta LDOS(\varepsilon)$, shown in Fig. \ref{fig:ldos}. This function indicates the changes in LDOS caused by a contact of two slabs, both spatially and energetically resolved. Analysing this function, we do not observe significant changes in the vicinity of band edges meaning that the charge transfer of electrons near band edges is negligibly small. Instead, we clearly see a non-negligible hybridization at certain energies. The in-plane panel in Fig. \ref{fig:ldos} illustrates the splitting of silicon atomic orbitals into bound and anti-bound states. Therefore, the dipole layer is caused by a weak hybridization of atomic orbitals and does not depend much on the band edge alignment. The most dramatic changes in $LDOS(\varepsilon)$ in the contacting molecular layer whereas $\Delta LDOS(\varepsilon)$ is close zero almost for all energies for the second molecular layer. 

The observed orbital hybridization can not be interpreted as a formation of covalent bonds between Si and Tc slabs, since both bound and antibound states are occupied. This effect is rather attributed to the so-called Pauli pushback effect described in Ref. \cite{Zojer} when the contact of two slabs leads to a redistribution of charge carriers minimizing the overlap between orbitals respecting the Pauli exclusion principle. As a result, we observe a reduction of the charge density at the contact of two slabs which is shown by blue color in Fig. \ref{fig:si_tc_bands2}c. This change is compensated by an increase of charge density mostly between first and second atomic layers of Si, marked by red color in Fig. \ref{fig:si_tc_bands2}c. This effect is seemed to be independent of DFT parameters and can not be associated with limitations of the predictive power of DFT.

The discussion on the dipole layer provides an important information that has to be considered prior to the GW computations. In the case when it is physical, we can use the simplified G$_0$W$_0$ approximation without self-consistent loop, assuming that DFT provides correct results on the redistribution of electrons induced by a contact of two slab. On the other hand, if the dipole layer is spurious, DFT results are neither reliable nor providing a good basis set for the G$_0$W$_0$ approximation. In principle, the error associated with such a spurious charge transfer can be eventually eliminated in the process of the fully self-consistent GW computations. It is desirable, however, to perform GW in a non-self-consistent way (single shot) to avoid the large computational cost of repeated self-consistent loops.

Throughout this work we compute projections of the Kohn-Sham orbitals on a rectangular box confining a single molecular layer (blue and red markers in Fig. \ref{fig:si_tc_bands1}). Due to weak van-der-Waals coupling between molecular layers, the projections take two values making it possible to classify energy bands according to their location in the first (closer to Si) or second molecular layer (surface layer). As a result, we can estimate band edges for each molecular layer. In the surface layer, according to our DFT computations, the band-gap is slightly larger and the valence and conduction band edges are shifted by about 50 and 100 meV respectively towards the vacuum level. As a result, the ionization energy is different for the contacting and surface layer which can be attributed to the studied previously orientation-dependent ionization energies in organic crystals.\cite{Duhm2008}

\section{Many-body perturbation theory results} 

\subsection{G$_0$W$_0$ approximation}

DFT usually significantly underestimates band-gap energy and ionization potentials. This discrepancy is related to an error associated with approximated exchange-correlation functionals, which is in turn related to a so-called delocalization error, also known as many-electron self-interaction error \cite{McKechnie, Cohen}. Better accuracy in modeling exchange-correlation effects can be achieved using MBPT in the GW approximation \cite{HEDIN19701, Onida, Golze}. Quasi-particle energies within MBPT in the GW approximation are given by the poles of the retarded Green's function $G$. Compared to the energies of the Kohn-Sham orbitals, they are renormalized by so-called GW corrections given by the  expectation values of the self-energy operator defined by Hedin's equations \cite{Onida,Golze}: 

\begin{equation}
    \delta \varepsilon_i = \bra{\phi_i} \Sigma\left[GW\right] \ket{\phi_i},
    \label{hedin1}
\end{equation}
where: $\phi_i$ is an element of the basis set (usually Kohn-Sham orbitals), $\Sigma\left[GW\right]$ is the self-energy matrix dependent on the retarded Green's function, $G$, and screened Coulomb potential, $W$. The latter reads:
\begin{equation}
    W = \epsilon^{-1} v = \left[1-v \chi \right]^{-1} v,
    \label{hedin2}
\end{equation}
where $\epsilon^{-1}$ is the dielectric matrix inverse, $v$ is the bare Coulomb potential, and $\chi$ is the irreducible polarizability matrix.

Since the system under consideration is rather large (348 atoms in the super-cell), it is desirable to avoid an iterative approach and use the non-iterative G$_0$W$_0$ approximation. The standard routine for this method implies the following sequence of computations: 1) compute the basis set represented by Kohn-Sham orbitals using DFT; 2) compute the irreducible polarizability, dielectric matrix and its  inverse in the random phase approximation\cite{PhysRevLett.45.290, PhysRevB.25.2867}, Eq. (\ref{hedin2}), using the basis set computed at the previous step; 3) compute the self-energy corrections to the DFT eigen-energies, Eq. (\ref{hedin1}), applying the first-principles methodology of Hybertsen and Louie with the generalized plasmon-pole model for the frequency-dependent dielectric matrix \cite{Louie}. 

For the first step, we use the results obtained in Sec.~II, keeping all the parameters unchanged. All the computations related to MBPT (the second and third steps) have been performed using BerkeleyGW \cite{BGW}. In ab-initio MBPT computations, it is important to ensure convergence of the results by properly choosing the kinetic energy cutoff for the dielectric matrix and the number of unoccupied bands participating in sums in both the dielectric matrix and Coulomb-hole self-energy. In the latter case, the convergence is rather slow (see the convergence tests in the Supplementary materials, Section 2). In this work we use two techniques to reduce the number of bands: first, we apply the modified static remainder approach \cite{Deslippe} and use the extrapolation technique based on fitting the Coulomb-hole self-energy by a hyperbolic function \cite{Tamblyn}. After a series of convergence tests discussed in Supplementary information we have derived the following parameters: the kinetic energy cutoff for the dielectric matrix is of 15 Ry, the sums run over 534 unoccupied orbitals (1200 orbitals in total), and the k-grid is same as for the DFT calculations discussed in Sec. II.

\begin{figure}[t]
    \centering
    \includegraphics[width=8.1cm]{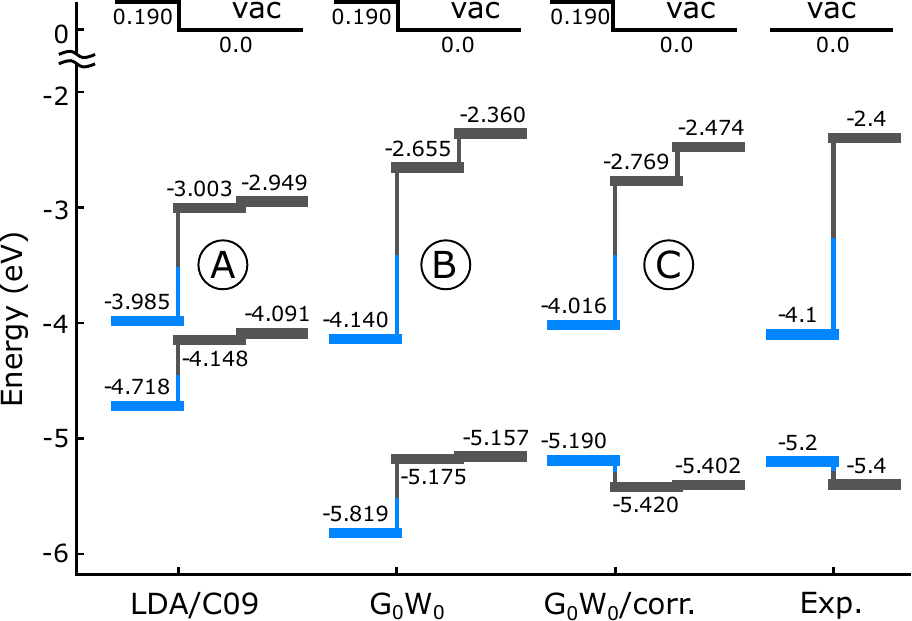}
    \caption{Band edge alignment at the Si/Tc interface for the supercell model obtained by DFT (diagram A); the G$_0$W$_0$ approximation (diagram B); a combination of the G$_0$W$_0$ approach, orbital relaxation correction for Tc and line-up potential model for Si (diagram C). The experimental results (the right-most diagram) are taken from Ref.~\cite{MacQueen}.}
    \label{fig:edges_supercell}
\end{figure}

The results of band edge computations for the supercell model are shown in Fig.~\ref{fig:edges_supercell}. Expectedly, the band structures obtained with MBPT have larger band gaps for both Tc and Si in comparison to those obtained by DFT (compare diagrams A and B in Fig. \ref{fig:edges_supercell}). What is somewhat surprising is that the shifts of the band edges in Si and Tc are qualitatively different. For Tc, the GW corrections to the valence and conduction band edges are more less the same, whereas for Si, the corrections substantially decrease the energy of the valence band edge while the conduction band edge remains almost unchanged. According to the results of Ref. \cite{Wei}, such behaviour for GW corrections for Si is the case for most of the inorganic III-V and IV semiconductors. Since the difference between DFT and GW results is determined by the exchange-correlation effects, we conclude that the nature of electron correlations is substantially different in Tc and Si slabs.

Comparing the results obtained by the G$_0$W$_0$ method (diagram B, Fig. \ref{fig:edges_supercell}) to experimental data, we find significant discrepancy: the energy of the valence band edge is largely underestimated for Si (about 0.6 eV) and the valence band edge for Tc is about 200 meV larger than the experimental value. In the case of Tc this discrepancy can be explained by the fact that we use the G$_0$W$_0$ approximation neglecting orbital relaxation. In the case of Si, due to largely delocalized wave functions, the source of the error is the slab model which is prone to the quantum confinement effect and is characterized by a modified electrostatic environment that reduces the dielectric screening for the G$_0$W$_0$ method.  This inaccuracy can be reduced using methods described in the next section.

\subsection{Orbital relaxation corrections for Tc slab}

The discrepancy between the G$_0$W$_0$ and GW approximations is attributed to the orbital relaxation due to rearrangement of electrons caused by the correlation effects. For the case of organic semiconductors with small orbit hybridization between molecules, we propose an approximation, according to which the orbital relaxation in isolated molecule is close to the orbital relaxation energy in the van-der-Waals molecular crystals. This approach is similar to the so-called QM/QM' method \cite{C9CP06154A} and leads to the following expressions for the band edges in molecular crystals:

\begin{equation}
    \varepsilon_j =  \varepsilon_{j}^{G_0W_0/crys} + \left(\varepsilon_{j}^{GW/mol} - \varepsilon_{j}^{G_0W_0/mol} \right),
    \label{eq:corr0}
\end{equation}
where: $j \in \{ IP, EA \}$, $\varepsilon_{j}^{G_0W_0/crys}$ is the band edge of the molecular crystal computed with the G$_0$W$_0$ approach, $\varepsilon_{j}^{G_0W_0/mol}$ and $\varepsilon_{j}^{GW/mol}$ are the orbital energies of the isolated molecule computed with the G$_0$W$_0$ and GW approaches respectively.

In predicting ionization potentials of molecules, the full-self-consistent GW method gives precision comparable to the so-called $\Delta$SCF approach\cite{McKechnie}. The latter, however, is less computational demanding. Therefore, we use the $\Delta SCF$ method instead of GW method for a single molecule:

\begin{equation}
    \varepsilon_j =  \varepsilon_{j}^{G_0W_0/crys} + \left(\varepsilon_{j}^{\Delta SCF} - \varepsilon_{j}^{G_0W_0/mol} \right),
    \label{eq:corr}
\end{equation}
where:
\begin{equation}
     \varepsilon_{j}^{\Delta SCF} = E^{mol}(N\pm 1) -E^{mol}(N).
     \label{eq:corr1}
\end{equation}

\begin{figure}[t]
    \centering
    \includegraphics[width=8.1cm]{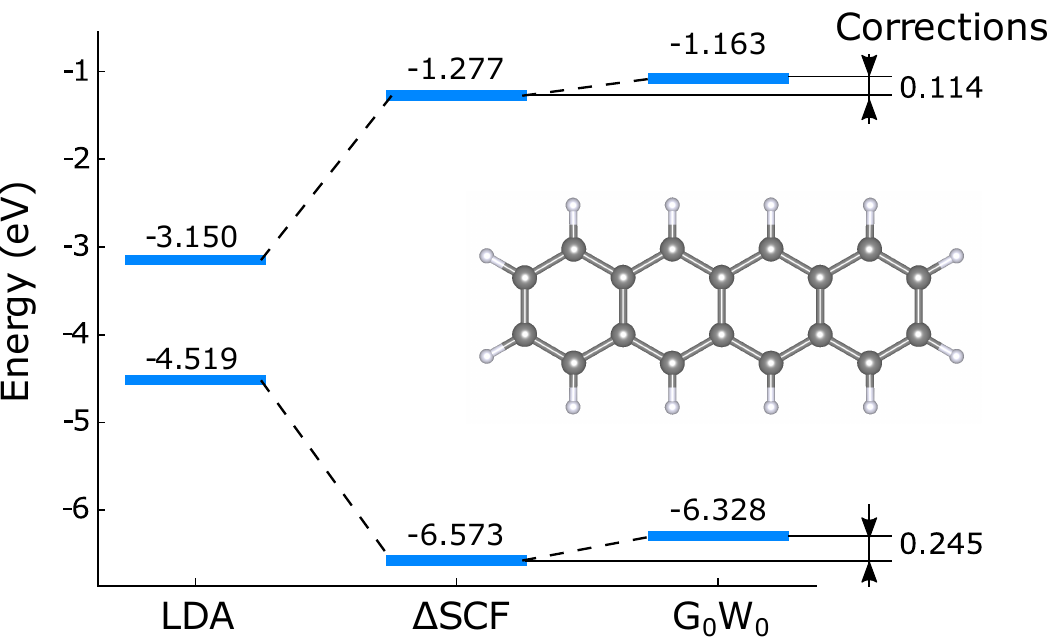}
    \caption{HOMO and LUMO energies of the isolated Tc molecule computed by different methods: density function theory with the local-density approximation (LDA), $\Delta$SCF method and MBPT with the G$_0$W$_0$ approximation. The difference between the results obtained by $\Delta$SCF method and results obtained by G$_0$W$_0$ methods is attributed to the orbital relaxation energy.}
    \label{sm_corr}
\end{figure}

The energy difference in the parentheses in Eq. (\ref{eq:corr}) is shown in Fig. \ref{sm_corr} where we compare different computational approaches used to compute HOMO and LUMO energies for a single Tc molecule. The difference between the results obtained by the LDA and $\Delta$SCF methods is the origin of Koopmans' corrections in DFT \cite{McKechnie}. This discrepancy is attributed to the so-called delocalization error or many-electron self-interaction error manifesting itself as an incorrect dependence of the total energy on the fractional number of electrons (convex instead of linear). The error is caused by neglecting the electron correlation effects and orbital relaxation. The G$_0$W$_0$ approach, in turn, takes into account the correlation effects neglecting the orbital relaxation \cite{McKechnie}. Therefore, the difference between the $\Delta$SCF and G$_0$W$_0$ energies can be attributed to the orbital relaxation effect only. These corrections for HOMO and LUMO are very close to each other and decrease the orbital energies as is expected after the orbital relaxation. Note that the atomic configuration of the Tc molecule for computing the energy corrections is taken unchanged from the Tc slab after the geometry optimization in the supercell. Also, all the DFT parameters, such as the exchange-correlation functional and pseudopotentials, should be the same for both G$_0$W$_0$ and $\Delta$SCF computation in order to ensure the mean-field contributions cancel each other after subtracting the results as is explained in details in the Supplementary materials (Sec. 4).

 This method is consistent with the recently proposed Wannier-Koopmans method \cite{Ma2016, Ma_Neaton}. The relationship between two approaches in the approximation of negligible orbital hybridization between organic molecules is discussed in the supplementary materials (Sec. 5).

\subsection{Line-up potential method for Si slab}

The valance band edge in Si obtained by the G$_0$W$_0$ approach (diagram B in Fig.  \ref{fig:edges_supercell}) is about 0.6 eV lower in comparison to the experimental data, while the position of the conduction band edge is accurate within 50 meV. Similar results are reported in Ref. \cite{Wei} where the ionization energy for the Si slab with the (111) 2x1 reconstructed surface is -5.64 eV which is pretty close to our results: -5.71 eV for the free surface and -5.618 eV for the surface in contact with the Tc bi-layer. The different types of surface termination (the (111)-surface against the (100)-surface) leads to the ionization potential change of about 0.12 eV \cite{PhysRevB.64.195305} and its effect is not important for this consideration. In all these cases, some of the discrepancies between the experimental data and the results of the computations can be attributed to the fact that the experimental data are reported for surfaces of the bulk semiconductors while the computations are performed for the slab models where the band edges are shifted due to the quantum confinement and truncated dielectric screening in slabs. In order to accurately simulate a bulk semiconductor, the slab thickness has to be about 5~nm which, together with the Tc slab, makes the system extremely large and practically intractable for most of the computational methods of quantum chemistry.

\begin{figure}
    \centering
    \includegraphics[width=8.1cm]{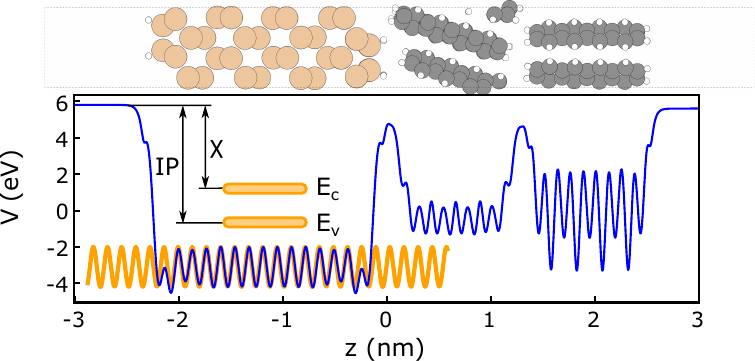}
    \caption{Electrostatic potential computed within DFT for the Si/Tc slab (blue line) and band edges $E_v$ and $E_c$ for bulk Si (orange thick lines) computed relative to the averaged electrostatic potential in the bulk Si unit cell (orange thin line) lined-up to the averaged electrostatic potential in the middle of the Si slab.}
    \label{fig:pot_bulk}
\end{figure}

This discrepancy, however, can be somewhat compensated using the line-up potential method \cite{Borriello, Marri}. This method relies on the idea that local physical quantities, such as the electrostatic potential, in the middle of the slab should coincide with the analogous quantities for the bulk configuration. The DFT computations for the bulk materials give the valance and conduction band edges measured relative to the averaged electrostatic potential (see Fig. \ref{fig:pot_bulk}). These quantities are not prone to the quantum confinement effect and reduced screening in the slab model. However, the computations for the slab model are still needed to estimate the vacuum level relative to the electrostatic potential and, correspondingly, the band edges in order to obtain the ionization potential and electron affinity: the band edges ordained from the bulk case can be properly aligned relative to the vacuum level of the slab by equating the averaged electrostatic potentials in the bulk semiconductor and slab (see Fig. \ref{fig:pot_bulk}).

\subsection{Effect of the corrections}

Adding the orbital relaxation corrections computed with Eq. (\ref{eq:corr}) and (\ref{eq:corr1}) and computing Si band edges using the line-up potential method improve agreement with experimental data (see diagram C in Fig. \ref{fig:edges_supercell}). This combination of the G$_0$W$_0$ approach, orbital relaxation correction and line-up potential method predicts a type-I heterostructure for the Si/Tc interface, as is experimentally observed. In absolute values, the contribution from the GW correction is the most pronounced, next comes the contribution from the line-up potential method for Si and the least significant is the orbital relaxation contribution for Tc.

The resulting corrections for Tc and Si are so different that they are capable of switching the type-II heterostructure (predicted by DFT) to the type-I (predicted by MBPT/GW). In inorganic heterostructures, the GW corrections are approximately the same for all semiconductors, so the band offsets predicted by DFT remain unchanged after GW corrections. Therefore, the case when the electron correlations determine the band offsets is specific for HIOS interfaces only and shows the extraordinary importance of the correlation effects for them.

The computed band offsets are 0.23 meV for the valence band and 1.247 eV for the conduction band for the contacting Tc layer. The next Tc layer has a slightly larger band gap and, correspondingly, slightly different band offsets, see Fig. \ref{fig:edges_supercell}c.

\subsection{Dielectric embedding approach and dielectric screening}

In order to analyse the interplay between the electrostatic field of the dipole layer and mutual dielectric screening of the slabs on energy band alignment, it is productive to isolate one effect from the other by comparing the results obtained from the supercell model with the results obtained by the dielectric embedding technique for the GW method \cite{Liu2019, Liu2020}. This technique is based on the idea that, if slabs are weakly coupled and the orbital hybridization between them can be neglected, the basis set $\phi_i$ in Eq. (\ref{hedin1}) and (\ref{hedin2}) can be split into two subsets of non-overlap functions for the Tc and Si slabs, $\phi_i^{Tc}$ and $\phi_i^{Si}$ and the Green's function in this representation can be computed independently for Si and Tc slabs, $G^{Si}$ and $G^{Tc}$. As a result, the GW self-energies read:

\begin{equation}
\begin{split}
    &\Sigma^{Si} \approx \bra{\phi_i^{Si}} \Sigma\left[G^{Si}W\right] \ket{\phi_i^{Si}}, \\
    &\Sigma^{Tc} \approx \bra{\phi_i^{Tc}} \Sigma\left[G^{Tc}W\right] \ket{\phi_i^{Tc}}.
\end{split}
\label{sigma}
\end{equation}

In this model, the slabs are coupled only via the screened potential $W$ with the mutual electrostatic effects and long-range electron-electron correlation quantitatively expressed by the frequency-dependent dielectric function $\epsilon^{-1}= \left[1-v \chi \right]^{-1}$ in Eq. (\ref{hedin2}). The dielectric embedding is based on the assumption that the polarizability matrix $\chi$ is additive and can be computed independently for each slab \cite{Liu2019, Liu2020}:
\begin{equation}
\chi \approx \chi^{Si} + \chi^{Tc}.
\label{chi}
\end{equation}
The bottleneck in the numerical computations is computing the matrix inverse. 

\begin{figure}
    \centering
    \subfigure[]{\includegraphics[width=3cm]{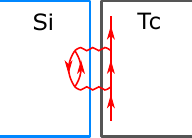}} \hspace{.3cm}
    \subfigure[]{\includegraphics[width=3cm]{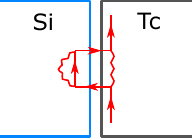}}
    \caption{Examples of Feynman diagrams representing the terms in the Green's function expansion involving a) exchanging of virtual photon and b) exchanging electrons  between slabs. In the weakly-coupled slabs model the diagrams of the second kind are neglected.}
    \label{fig:feynmann}
\end{figure}

 Originally, this idea has been proposed to speed up the GW computations reducing sizes of the super-cell when computing $G$, $\chi$ and $\Sigma$ for organic molecules weakly coupled to metallic substrates \cite{Liu2019, Liu2020}. The dielectric embedding technique implies insulating slabs electronically neglecting all the propagators that involve particle propagation between organic and inorganic semiconductors, keeping only diagrams where two semiconductors are coupled by exchanging virtual photons (see Fig. \ref{fig:feynmann}). This follows from Eq. (\ref{sigma}) where the self-energy is expressed in terms of Green's functions for which both initial and final states belong either to Tc or to Si. In the dielectric embedding, the coupling between slabs is included into consideration only via the screened Coulomb potential given by Eqs. (\ref{chi}) and (\ref{hedin2}).

\begin{figure*}[t]
    \centering
    \includegraphics[width=0.8\linewidth]{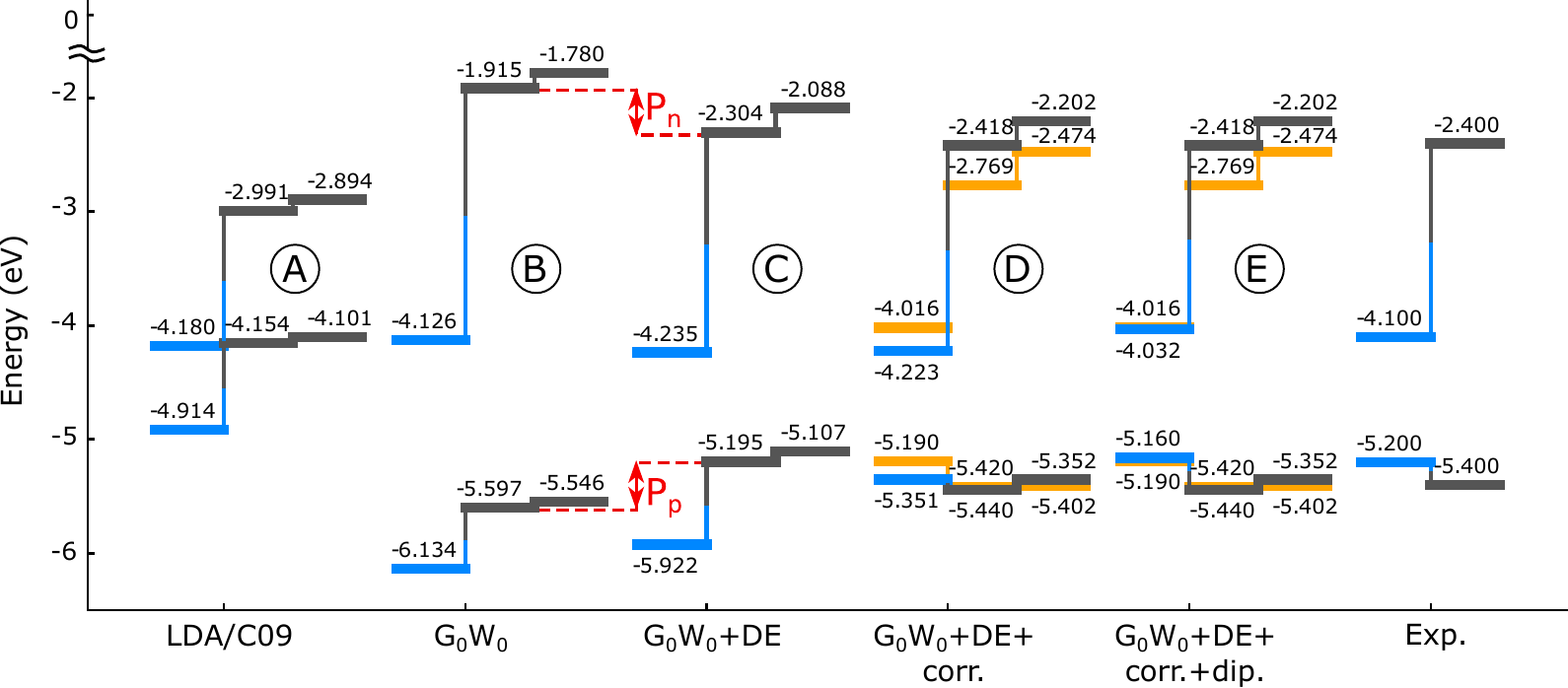}
    \caption{Band edge alignment for the isolated Si and Tc slabs computed within DFT (diagram A) and G$_0$W$_0$ approach (diagram B). Diagrams C-E represent the band edge alignment computed using G$_0$W$_0$ approximation with the dielectric embedding approach without corrections (diagram C), with corrections (diagram D) and with combined corrections and dipole layer potential (diagram E). The corrections are given by the line-up potential model for Si and orbital relaxation correction for Tc. The orange color denotes the band edges obtained from the supercell model, diagram C in Fig. \ref{fig:edges_supercell}.  The experimental results  (the right-most diagram) are taken from Ref.~\cite{MacQueen}.}
    \label{fig:edges_de}
\end{figure*}

The DFT and MBPT results for the isolated slabs as well as the results with dielectric embedding are shown in Fig. \ref{fig:edges_de}. The comparison between the GW results for isolated slabs (diagram B in Fig. \ref{fig:edges_de}) and the result computed with the dielectric embedding (diagram C in Fig. \ref{fig:edges_de}) reveals the contribution from the mutual dielectric screening between slabs to the energy band alignment. Quantitatively these contributions can be expressed in terms of the polarization  energies:
\begin{equation}
\begin{split}
    &P_p=E_v^{G_0W_0+DE}-E_v^{G_0W_0},\\   
    &P_n=E_c^{G_0W_0+DE}-E_c^{G_0W_0}.
\end{split}
\end{equation}

For the contacting Tc layer, $P_p = 0.389$ eV and $P_n = 0.402$ eV, which is almost twice large the energy of the dipole layer. The dielectric embedding leads to smaller band gaps in both materials, however these changes are more pronounced for the organic semiconductor and for the first molecular layer. This effect is similar to the results of previous studies of the dependence of the frontier orbitals of a single organic molecule on the distance to a metallic substrate \cite{Neaton1}.

After computing the GW corrections with the dielectric embedding, we apply the orbital relaxation corrections for the organic semiconductor and the line-up potential method for the inorganic semiconductor as we did previously for the supercell model. The resulting band edges, shown in diagram D in Fig. \ref{fig:edges_de}, are characterized by a significantly smaller offset in the valance band compared to the supercell model. This is because the dielectric embedding approach does not take into account electron redistribution caused by the contact of two slabs and the corresponding dipole layer. We can use the energy of the dipole layer obtained from the DFT calculations and shift the band edges accordingly (see diagram E in Fig. \ref{fig:edges_de}). However, even in this case, we observe some discrepancy in predicting the band gaps of Tc and conduction band offsets, especially for the contacting Tc layer. This discrepancy can be explained by a perturbation of the electronic structure because of the slight hybridizing of molecular orbitals with silicon.

\section{Discussion and conclusions}

The results of this work have demonstrated the exceptional importance of the exchange-correlation effects in modelling the energy band alignment for the HIOS interfaces. For the Tc/Si interface, the energy band alignment is determined by an interplay of the mutual dynamic dielectric screening of two materials and the formation of a dipole layer due to the Pauli pushback effect \cite{Zojer}. Using the dielectric embedding technique, we were able to isolate the effect of the mutual dielectric screening from the effect of the dipole layer in order to estimate the contribution of each of them separately. For the van-der-Waals interface considered in this work, the contribution from the mutual screening, determined by the computed polarization energies $P_p$ and $P_n$, is several times larger than the dipole layer energy $E_{dip}$. 

This result have practical implications. Conventionally, the energy band alignment can be engineered by functionalizing Si with acceptor or donor molecules that modify the work function of the surface and the dipole layer \cite{Niederhausen2020}. Additionally to this approach, we have demonstrated that the dielectric screening produced by substrate also contributes to the energy band alignment. Therefore, modifying the dielectric properties of the substrate (for instance by using the high-k dielectrics) can modify the energy band edges of the contacting organic layer up to 0.5 eV.  

The mean-field theories are not able to predict the energy band offsets for the HIOS interfaces even qualitatively. This fact has been well illustrated by computing GW corrections for organic and inorganic parts of the Tc/Si heterostructure. Unlike in the inorganic semiconductors, these corrections change across the interface dramatically (several hundreds meV), thus substantially contributing to the band offsets.

Besides computational challenges related to the exchange-correlation effect,  supercell models with slabs representing an interface face problems related to the reduced dielectric screening and quantum confinement effects that can affect the accuracy of the results, especially for the inorganic part. We have demonstrated that this problem can be tackled using the line-up potential method for the HIOS interfaces. Also, the use of the non-self-consistent G$_0$W$_0$ approximation for the organic part has some inaccuracy (up to 300 meV) which can be compensated by the orbital relaxation corrections proposed in this work.

\section{Acknowledgements}
The authors acknowledge the support of the Australian Research Council through grant CE170100026. Work at the Molecular Foundry was supported by the Office of Science, Office of Basic Energy Sciences of the U.S. Department of Energy under Contract No. DE-AC02-05CH11231. This project was undertaken with the assistance of resources and services from the National Computational Infrastructure (NCI), which is supported by the Australian Government. This research used resources of the National Energy Research Scientific Computing Center, a DOE Office of Science User Facility supported by the Office of Science of the U.S. Department of Energy under Contract No. DE-AC02-05CH11231.

\section{Supporting Information}

The supporting information contains the results of convergence tests aimed to determine DFT parameters and parameters for GW computations, the band structure obtained within DFT for the supercell model and for isolated Tc and Si slabs and the description of the relationship between the orbital relaxation corrections and Wannier-Koopmans method. 

\bibliographystyle{MSP}
\bibliography{tc_si}

\end{document}